\documentclass[]{aa}
\usepackage{graphicx}
\usepackage{txfonts}
\def\OII{[O{\tt II}]$\lambda$3727}
\def\oii{[O{\tt II}]}
\def\OIII{[O{\tt III}]$\lambda$5007}
\def\Halp{H$\alpha$+[N{\tt II}]$\lambda\lambda$6548,6583}
\def\Hbeta{H$\beta$}
\def\D4{$D4000$}
\def\uvcol{\hbox{$C$(28-39)}}
\begin{document}
   \title{The K20 survey.}

   \subtitle{VII. The spectroscopic catalogue: spectral
   properties and evolution of the galaxy population
   \thanks{Based on observation mad at the European Southern
           Observatory, Paranal, Chile (ESO LP 164.O-0560)}$^,$
   \thanks{Table~\ref{tabcat} is also available in electronic form
      at the CDS via anonymous ftp to cdsarc.u-strasbg.fr (130.79.128.5)
    or via http://cdsweb.u-strasbg.fr/cgi-bin/qcat?J/A+A/}
   }
   \author{M.~Mignoli
          \inst{1}
          \and
          A.~Cimatti
          \inst{2}
          \and
          G.~Zamorani
          \inst{1}
          \and
          L.~Pozzetti
          \inst{1}
          \and
          E.~Daddi
          \inst{3}
          \and
          A.~Renzini
          \inst{3}
          \and
          T.~Broadhurst
          \inst{4}
          \and
          S.~Cristiani
          \inst{5}
          \and
          S.~D'Odorico
          \inst{3}
          \and
          A.~Fontana
          \inst{6}
          \and
          E.~Giallongo
          \inst{6}
          \and
          R.~Gilmozzi
          \inst{3}
          \and
          N.~Menci
          \inst{6}
          \and
          P.~Saracco
          \inst{7}
          }
   \offprints{M.~Mignoli \\
   \email {marco.mignoli@bo.astro.it}}
   \authorrunning{M.~Mignoli et.al}

   \institute{Istituto Nazionale di Astrofisica, Osservatorio Astronomico
              di Bologna, via Ranzani 1, 40127 Bologna, Italy
         \and
              Istituto Nazionale di Astrofisica, Osservatorio Astrofisico
              di Arcetri, Largo E.Fermi 5, 50125 Firenze, Italy
         \and
              European Southern Observatory, Karl-Schwarzschild-Str.~2,
              85748 Garching, Germany
         \and
              Racah Institute for Physics, 
              The Hebrew University, Jerusalem 91904, Israel
         \and
              Istituto Nazionale di Astrofisica, Osservatorio Astronomico
              di Trieste, via Tiepolo 11, 34131 Trieste, Italy
         \and
              Istituto Nazionale di Astrofisica, Osservatorio Astronomico
              di Roma, via dell'Osservatorio~2, Monteporzio, Italy
         \and
              Istituto Nazionale di Astrofisica, Osservatorio Astronomico
              di Brera, via E.~Bianchi~46, Merate, Italy
             }

   \date{Received September .....; accepted .....}

   \abstract{The K20 survey is a near infrared-selected, deep ($K_s < 20$)
   redshift survey targeting galaxies 
   in two independent regions of the sky, the Chandra Deep Field South and
   the field around the quasar $0055-2659$, for a total area of 52 arcmin$^2$.
   The total $K_s$-selected sample includes 545 objects. 
   Low-resolution ($R \approx 300-600$) optical spectra for 525 of them
   have been obtained with the FORS1/FORS2 spectrographs at the ESO/VLT,
   providing 501 spectroscopic identifications
   (including 12 type-1 AGN and 45 stars);
   consequently, we were able to measure redshifts and identify stars in
   96\% of the observed objects, whereas the spectroscopic completeness with
   respect to the total photometrically selected sample is 92\% (501/545).
   The K20 survey is therefore the most complete spectroscopic survey
   of a near~infrared-selected sample to date.
   The K20 survey contains 444 spectroscopically identified galaxies,
   covering a redshift range of $0.05 < z < 2.73$, with a mean redshift
   $<z>=0.75$; excluding the 32 ``low-quality'' redshifts does not
   significantly change these values.
   %
   %
   This paper describes the final K20 spectroscopic catalogue, along with
   the technique used to determine redshifts, measure the spectral
   features and characterize the spectra. The classification of the galaxy
   spectra has been performed according to a simple parametric recipe that
   uses the equivalent widths of the two main emission lines
   (\OII \ and H$\alpha$+[N{\tt II}]) and two continuum indices
   (the 4000\AA \ break index, \D4, and a near-UV color index, \uvcol).
   We defined three main spectroscopic classes: red early-type galaxies,
   blue emission-line galaxies and the intermediate galaxies,
   which show emission lines but a red continuum. More than 95\% of
   the examined galaxies is included in one of these spectral types and
   a composite spectrum is built for each of the three galaxy classes.
   The full spectroscopic catalogue, the reduced individual spectra and
   the composite spectra are released to the community through
   the K20 web page ({\tt http://www.arcetri.astro.it/$^\sim$k20/}).
   
   The blue emission-line and the early-type galaxies
   have been divided in redshift bins, and the corresponding composite
   spectra have been built, in order to investigate the evolution
   of the spectral properties of the K20 galaxies with redshift.
   The early-type average spectra are remarkable in their
   similarity, showing only subtle but systematic differences in
   the \D4 index, which are consistent with the ageing of the stellar
   population. Conversely, the star-forming galaxies present
   a significant ``blueing'' of the optical/near-UV continuum
   with redshift, although the \oii \ equivalent width remains
   constant ($\sim 33$\AA) in the same redshift intervals.
   We reproduce the observed properties with simple, dust-free population
   synthesis models, suggesting that the high-redshift galaxies are
   younger and more active than those detected at lower redshift,
   whilst the equivalent width of the emission lines apparently
   require a lower metallicity for the low-redshift objects. 
   This may be consistent with the metallicity-luminosity
   relationship locally observed for star-forming galaxies.
   
   }

   \maketitle
%

\section{Introduction}

Despite the success of photometric techniques in 
selecting samples of high-redshift galaxies, spectroscopy still 
represents a more powerful approach to investigate the nature and 
evolution of galaxies thanks to the larger amount of information 
contained in galaxy spectra with respect to broad-band photometry. 
The 8--10m-class telescopes and their modern multi--object spectrographs 
allow to perform deep spectroscopic surveys and derive the spectral 
properties of large samples of faint galaxies.

Among the several tools currently available to investigate 
galaxy evolution, the selection and study of samples of faint 
galaxies in the near-infrared (especially in the $K$-band) is one 
of the most powerful for ground-based observations thanks to three 
main features. First, the $K$-band selection is free from strong biases
against or in favor of particular classes of galaxies and minimizes the
k-correction effects up to redshift $\sim 3$ (Cowie et~al. \cite{HawaiiK}). 
Second, the effects of dust extinction and obscuration are smaller
in near-infrared samples. Third, as the rest-frame near-IR 
luminosity is known to trace the galaxy mass, samples
of galaxies selected in the $K$-band allow to trace the cosmic
history of galaxy mass assembly up to $z\sim2$. These advantages
make $K$-band surveys, although they contain a smaller number of
objects than the optical ones, more suitable to investigate
galaxy evolution with respect to optical samples, the latter being
biased toward star-forming galaxies with little dust extinction
and sensitive to the rest-frame ultraviolet luminosity dominated
by hot short-lived massive stars (i.e., not representative
of the global galaxy stellar mass).

Following this approach, several near-infrared surveys have been
and are being undertaken, with some of them based mostly or exclusively
on photometric redshifts and others making use of deep spectroscopy
(see Salvato et~al. 2004\footnote{http://www.mpe.mpg.de/$^\sim$mara/surveys/}
for a recent "survey of surveys"). 
Among the optical spectroscopic surveys designed to identify 
faint $K$-selected galaxies down to $K_s \sim20-20.6$, there are those of
Songaila et~al. (\cite{songaila}), Cowie et~al. (\cite{cowie96}),
Cohen et~al. (\cite{CFGRS_7}, \cite{CFGRS_8}; CFGRS), 
Drory et~al. (\cite{MUNICS}; MUNICS), Cimatti et~al.
(\cite{K20_3}; K20), and Abraham et~al. (\cite{GDDS}; GDDS).
The typical size of the spectroscopic samples is in the range of
about 200--500 objects selected from fields of 15--100 arcmin$^2$.
The percentages of objects with spectroscopic identification 
with respect to the photometric samples range from $\sim$80\%
(e.g. Cohen et~al. \cite{CFGRS_8}; Abraham et~al. \cite{GDDS})
to $\sim$90\% (e.g. Cimatti et~al. \cite{K20_3};
this work). The two most recent spectroscopic surveys (K20 and GDDS) were
designed with two different approaches. The GDDS sample was selected to
$K_s<20.6$, covers an area of 30$\times$4 arcmin$^2$, and very deep
spectroscopy  was concentrated on a subset of candidates at $z>0.8$.
The K20 survey was based on a sample selected at $K_s<20$ from an area
of 52 arcmin$^2$, and spectroscopy was carried out for all the objects
without any redshift pre-selection; the spectroscopic sample contains
444 galaxies covering a wide redshift range of $0.05 < z < 2.73$,
with a mean redshift $<z>=0.75$.
Both surveys were capable of covering the so-called ``redshift desert'' by
identifying a substantial number of galaxies at $1.4<z<2.7$. 

The importance of near-infrared surveys has been demonstrated 
by the discovery of substantial populations of distant galaxies which 
are missed by traditional optical surveys (see McCarthy \cite{McCarthy}
for a recent review), and by the possibility of using the resulting samples
to successfully investigate several aspects of galaxy evolution
(e.g. Cimatti \cite{Cimatti04}), to perform stringent comparisons with models
of galaxy formation (e.g. Cimatti et~al. \cite{K20_4}; Somerville et~al. 
\cite{Somerville04}), and to place the first constraints on the evolution
of the stellar mass density in the Universe (e.g. Dickinson et~al.
\cite{dickinson};  Drory et~al. \cite{drory}; Fontana et~al. \cite{K20_6};
Glazebrook et~al. \cite{glaze}). Overall, the results of near-infrared
surveys suggest a scenario where the luminosity function, and the stellar
mass density and function are characterized by a little and slow
evolution to $z\sim1$. The picture becomes more controversial at higher
redshifts, where the evolution of the near-IR luminosity function
and cosmic stellar mass density is faster, and the properties
and role of the massive galaxies found at $1.5<z<3$ (e.g. Franx et~al. 
\cite{franx}; van Dokkum et~al. \cite{vandokkum}; Daddi et~al. \cite{daddi2};
Glazebrook et~al. \cite{glaze}; Cimatti et~al. \cite{oldyoung};
Saracco et~al. \cite{saracco}) have still to be clarified. 

In this paper, we present the spectroscopic sample resulting from the
K20 survey, discuss the spectral classification and main properties
of the identified objects, show the average spectra of different
classes of galaxies, and explore the existence of spectral evolutionary
trends as a function of redshift. Together with this paper, we publicly 
release the K20 sample, the individual spectra and the average spectra
on the K20 survey web page. $H_0=70~$kms$^{-1}$Mpc$^{-1}$, $\Omega_m=0.3$ 
and $\Omega_{\Lambda}=0.7$ are adopted throughout this paper.

\section{The K20 survey and its main results}

The K20 survey was based on an ESO VLT Large Program aimed at obtaining
mostly optical and some near-infrared spectra of a complete sample of
545 objects down to $K_s \le 20.0$ (Vega system), to derive their
redshifts and spectral properties, and to investigate galaxy evolution
up to $z \sim 2$. 

\begin{figure}[t]
 \centering
  \resizebox{\hsize}{!}{\includegraphics{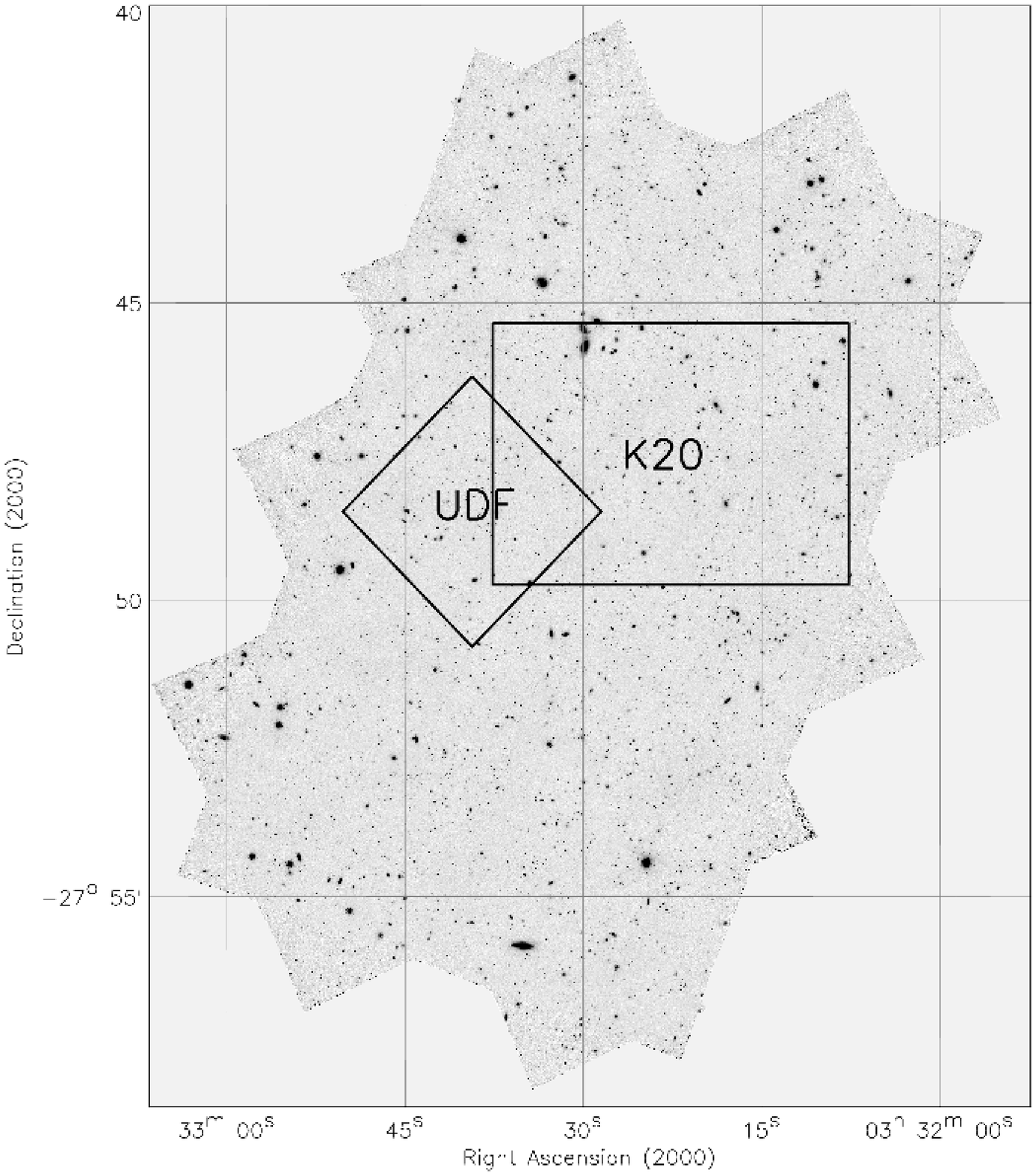}}
    \caption{\footnotesize
   {The K20 (rectangle, centered at $03^h32^m22\,\fs5$~~-$27\degr46'23\farcs5$,
   J2000) and the Hubble Ultra Deep (square, centered at
   $03^h32^m39\,\fs0$~~-$27\degr47'29\farcs1$) fields in the CDFS region,
   superimposed on the GOODS HST/ACS z-band mosaic.}
    }
   \label{figk20_hudf}
\end{figure}

The sample was extracted from two independent fields, with a total surveyed 
area of 52 arcmin$^2$. The targets were selected from a 32.2 arcmin$^2$ 
area of the Chandra Deep Field South (CDFS; Giacconi et~al. \cite{CDFS}) 
and from a 19.8 arcmin$^2$ area around the z=3.7 quasar $0055-2659$ (Q0055; 
Cristiani et~al. \cite{Q0055}).
The K20 32.2 arcmin$^2$ field in the CDFS is also located
within the 150 arcmin$^2$ southern field of the GOODS project, where
several multi-wavelength public data are or will be available (Dickinson
et~al. \cite{GOODS_1}; Giavalisco et~al. \cite{GOODS_2}). The K20 field 
in the CDFS overlaps marginally with the {\sl Hubble Ultra Deep Field}
(HUDF; Beckwith et al. 2005, in preparation; see Figure~\ref{figk20_hudf}).
The K20 sample characteristics, the photometric properties and completeness
of the sample, and the photometric redshifts were already described in
detail by Cimatti et~al. (\cite{K20_3}, hereafter Paper III). 

With respect to the spectroscopy discussed in Paper III (observations in 
1999-2000), additional spectroscopic data were obtained in 2002
by K20 team members (see Cimatti et~al. \cite{manifold}; Cimatti et al.
\cite{oldyoung}; Daddi et~al. \cite{daddi2}). 
This new spectroscopy allowed us to increase the overall spectroscopic 
redshift completeness of the total K20 sample (i.e. both K20 fields 
together) from 87\% (Paper III) to 92\%. Moreover, if also the ESO/GOODS
public spectra in the GOODS-South field published so far (Vanzella et~al.
\cite{fors2goods}) are added\footnote{For five galaxies the redshift
has been obtained from ESO/GOODS data.}, the spectroscopic redshift
completeness in the CDFS/K20 32.2 arcmin$^2$ sub-field alone is 94\%.
We emphasize that the K20 survey represents a 
significant improvement with respect to other surveys for faint 
$K$-selected galaxies thanks to its high spectroscopic redshift 
completeness (the highest to date) extended to very faint red objects, 
the larger sample and the availability of well-trained and reliable
photometric redshifts (see Fig.~9 and 10 of Cimatti et~al. \cite{K20_3}).

The main scientific results of the K20 survey have been published in 
a number of recent papers which discuss various aspects of galaxy 
evolution: the spectral and clustering properties of Extremely Red Objects
(EROs) (Cimatti et~al. \cite{K20_1}, Paper I; Daddi et~al. \cite{K20_2},
Paper II), the comparison between the observed redshift distribution
and those predicted by different models of galaxy formation (Cimatti
et~al. \cite{K20_4}, Paper IV), the evolution of the near-infrared luminosity
density and luminosity (Pozzetti et~al. \cite{K20_5}, Paper V),
the X-ray and morphological properties of EROs (Brusa et~al. \cite{EROS_X}; 
Cimatti et~al. \cite{manifold}), the study of the large scale structure
in the CDFS (Gilli et~al. \cite{gilli}), the evolution of the stellar mass
density and function (Fontana et~al. \cite{K20_6}, Paper VI), 
the evolution of the galaxy blue Luminosity Function (Poli et~al. \cite{poli}),
the nature and role of massive star-forming galaxies at $z\sim2$
(Daddi et~al. \cite{daddi2}; de Mello et~al. \cite{demello}),
the discovery of old and massive spheroids up to $z\sim1.9$
(Cimatti et~al. \cite{oldyoung}),
the definition of a new photometric technique for a joint selection
of star-forming and passive galaxies at $z>1.4$ (Daddi et~al. \cite{culling}),
and the study of the rest-frame $B$-band morphology as a function of redshift
based on HST/ACS imaging of the CDFS field (Cassata et~al. \cite{cassata}).

The sample presented in this paper is formed by all the extra-galactic 
objects with a redshift measured from optical (or near-infrared)
spectra obtained by our team in the framework of the K20 survey:  
the total sample includes 545 objects, 347 in the CDFS 
and 198 in the Q0055 field. We obtained optical spectra for 525 
objects and identified 501 of them, including 45 stars. 
In addition, it was necessary to apply a further redshift restriction
($z\leq 1.6$) in order to perform the spectral classification,
since the wavelength range with the spectral features
used for classification cannot be seen in 
our optical spectra at higher redshifts.
Some relevant numbers are given in Table~\ref{tabnum}. 
%
   \begin{table}[t]
      \caption[]{K20 relevant numbers.}
         \label{tabnum}
     $$ 
         \begin{tabular}{lrrr}
            \hline
            \noalign{\smallskip}
            & Total &  CDFS & Q0055 \\
            \noalign{\smallskip}
            \hline
            \noalign{\smallskip}
            Full sample   & 545 & 347 & 198 \\
	    Observed      & 525 & 336 & 189 \\
	    Identified    & 501 & 325 & 176 \\
	    Stars         & ~45 & ~36 & ~~9 \\
	    AGN (type~1)  & ~12 & ~~8 & ~~4 \\
            Galaxies      & 444 & 281 & 163 \\
	    Galaxies with $zq=1^{(\mathrm{a})}$& 412 & 264 & 148 \\
	    Galaxies with $zq=1$ and $z\leq1.6$& 402 & 256 & 146 \\
             \noalign{\smallskip}
            \hline
         \end{tabular}
     $$ 
\begin{list}{}{}
\item[($^{\mathrm{a}}$)]$zq$ stands for `z-quality':
galaxies with secure redshifts have \hbox{$zq=1$}.
\end{list}
   \end{table}

\section{Spectroscopic observations and data reduction}

   The spectroscopic observations were obtained at the ESO Very Large
   Telescope (VLT) UT1 and UT2 equipped with FORS1 (in MOS mode, using
   19 movable slitlets) and FORS2 (in MXU mode, where up to $\sim$~50
   targets could be simultaneously observed through a laser-cut
   mask), respectively. The optical spectroscopy was collected during
   20 nights distributed over a period of four years (1999-2000, 2002),
   and different set-ups were adopted: in particular, a slit width of
   $\sim$1~arcsec and the grisms 150I, 200I and 300I were used, providing
   dispersion of 5.5, 3.9, 2.6~\AA/pixel and spectral
   resolution of R=260, 380, 660 respectively\footnote{See the ESO web page
   http://www.eso.org/instruments/fors/inst/grisms.html for more details.}.
   The integration times were generally comprised between 15 minutes
   and 8 hours, depending on the brightness of the sources. Dithering along
   the slits between two fixed positions was adopted, whenever possible,
   for the spectroscopy of faint objects in order to efficiently remove the
   CCD fringing and the strong sky lines in the red part of the spectra.

   The data reduction was carried out using IRAF\footnote{IRAF
   is distributed by the National Optical Astronomy Observatories, which
   are operated by the Association of Universities for Research in 
   Astronomy, Inc., under cooperative agreement with the National Science
   Foundation.} tasks and treating separately each slit in the masks.
   The frames were bias-subtracted and flat-field corrected in the standard
   way. Cosmic ray removal was performed on the individual images using
   the IRAF script {\tt xzap} from the DIMSUM package\footnote{Deep Infrared
   Mosaicking Software, a package written by Eisenhardt, Dickinson, Stanford
   and Ward, available at http://iraf.noao.edu/contrib/dimsumV2.}. For the
   non-dithered spectra, the sky background was removed using a third-order
   polynomial fit along the spatial direction and avoiding the region
   including the object. When the dithering technique was adopted, the
   individual exposures were co-added following the classical A-B-B-A 
   procedure, routinely used in near-infrared spectroscopy. As the objects
   were spatially well sampled, the shift-and-add operation was performed
   on integer number of pixel basis. After sky-subtraction, all the 
   spectra were optimally extracted (Horne \cite{optextr}).
   The wavelength scale calibration was achieved using the He-Ne-Ar arc
   spectra, obtained with the same mask configuration as the science 
   observations. These calibration spectra were used to obtain a polynomial 
   dispersion relation with typical r.m.s. values of the order of
   $\sim 0.5$~\AA.

   A precise absolute flux calibration is a difficult task for spectra
   accumulated over many nights, with different instrumental set-ups and
   in varying observing conditions. Only a relative flux calibration was 
   carried out using various standard stars for each run and a correction
   for the atmospheric extinction was applied using the ESO standard
   extinction curve. For most of the objects the relative flux calibration
   appears to be quite good, as demonstrated by the fact that repeated
   spectra of the same objects agree well with each other, although obtained
   with different slit/grism. However, we did not try to correct for the slit
   losses and as a consequence we do not have absolute measurements of the
   spectral fluxes. Therefore, the present study is mostly focused on flux
   ratio measurements (line equivalent widths and continuum indices),
   which are insensitive to absolute flux calibration uncertainties.
   
   Finally, averaging all the useful spectra obtained in each single MOS/MXU
   exposure, we created a template for the optimal correction of the
   atmospheric absorption bands around 6870~\AA \ (B-band) and 7600~\AA \
   (A-band) for each set of observations. These templates were obtained
   by fitting a low-order polynomial to the mean spectrum, dividing the
   average by this smooth continuum, and setting the correction template to
   the value of this ratio in the telluric band regions, and to unity
   elsewhere.

\section{Redshift determination and spectral features measurement}

   The majority of K20 objects was very faint with respect to the 
   background, and most of the spectra had rather low signal-to-noise
   ratio ($S/N$).
   Thus, the redshifts have been
   interactively measured using IRAF tasks, this approach being better
   at finding real spectral features than a completely automatic
   identification technique. Moreover, through the visual inspection
   of the two-dimensional sky-subtracted spectral image it is possible
   to check the reliability of the identified features.

\begin{figure}[t]
 \centering
  \resizebox{\hsize}{!}{\includegraphics{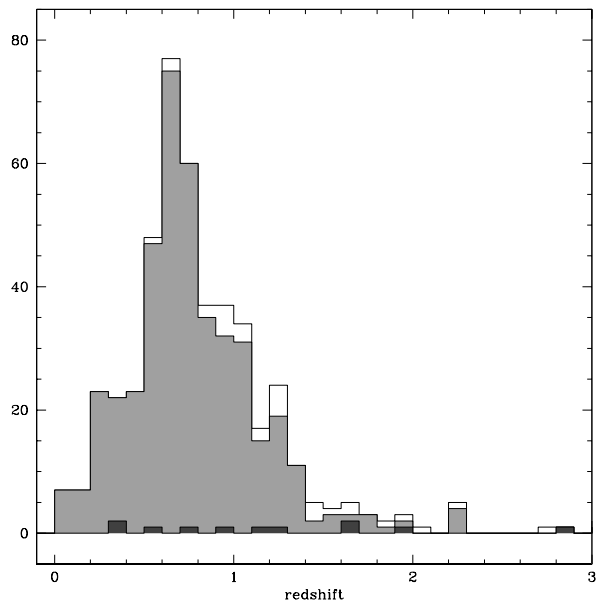}}
    \caption{\footnotesize
     The redshift distribution of the K20 spectroscopic catalogue for
     the whole extragalactic sample (empty histogram), the galaxies
     with ``secure'' redshifts ($zq=1$; light gray histogram) and
     the type~1 AGNs (dark gray histogram).}
   \label{fighistz}

\end{figure}

   The redshifts derived from absorption line spectra have been
   obtained using {\tt xcsao}, by
   cross-correlating the galaxy spectrum with a set of templates, and
   following Tonry \& Davis (\cite{xcorr}). In a first step we used as
   templates the spectra from Kinney et~al. (\cite{kinney}), then,
   to refine the redshift measurement, we adopted as template also
   the high-$S/N$ composite spectra obtained from our own
   survey database (see below). 
   For emission-line spectra, the redshift was obtained
   with the IRAF task {\tt rvidlines}, by means of multiple Gaussian
   fitting of the spectral features.
   All the redshifts were independently identified by two of the authors
   (MM, AC or ED), cross-checked, and the results were successively
   validated by at least one other member of the team. At the
   end of the analysis, the spectroscopic redshifts were divided into two
   categories: ``secure'' redshifts (when several features were detected
   with good confidence) and ``lower-quality'' redshifts 
   (when only one weak emission-line was detected and ascribed to 
   \OII, or when spectral features were only tentatively identified). 
   Galaxies with ``lower-quality'' flag ($zq = 0$) are 32, with
   respect to the total of 444 galaxies identified in the sample.
   The redshift distribution of the K20 spectroscopic catalogue
   is presented in Figure~\ref{fighistz} for both the galaxy and
   quasar samples. As shown in the figure, the average redshift
   of the galaxies with $zq = 0$ ($<z>=1.06$) is higher than that of
   the total sample.

   During the phase of redshift determination, also a ``visual'' (and
   somewhat subjective) spectral classification was attributed
   to each extragalactic object with secure redshift. 
   This preliminary spectral classification subdivided the
   objects into: type-1 AGN (i.e. with broad emission lines),
   emission-line galaxies, early-type galaxies with no emission lines
   and intermediate galaxies (red, early-type galaxy continuum \emph{and}
   emission lines).
   
\begin{figure}[t]
 \centering
  \resizebox{\hsize}{!}{\includegraphics{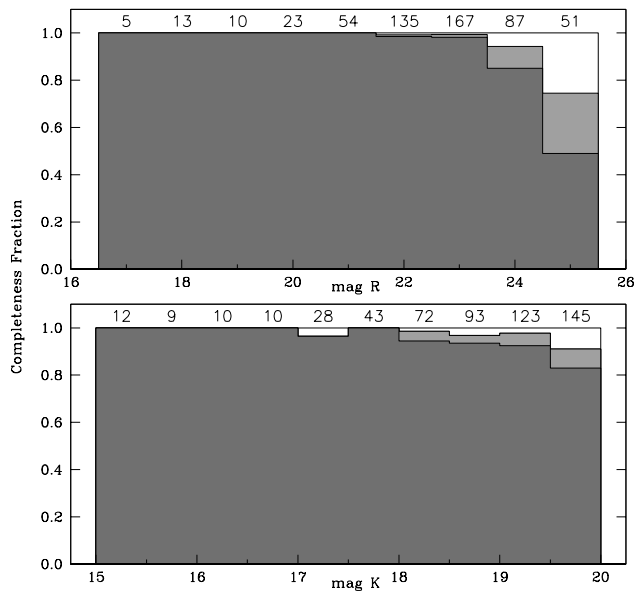}}
    \caption{\footnotesize
     Redshift measurement completeness in the K20 survey.
     The fractions as a function of magnitude of the observed targets 
     (light gray histogram) and of the objects with a redshift
     measurement (dark gray histogram) are shown both in
     the R band (top panel) and K band (bottom panel).
     The total number of sources in
     each magnitude bin is labeled along the top. The overall
     redshift measurement completeness is 92\%, whilst redshifts
     are measured for 95\% of the observed objects.}
   \label{figcompl}

\end{figure}

   The completeness of the spectroscopic identifications as a function
   of both $R$ and $K$ observed magnitudes is shown in Figure~\ref{figcompl}.
   Of the observed sample, 97\% is identified down to $K$=19.5;
   this efficiency drops to 91\% only in the faintest half-magnitude bin.
   As expected, and supported by Figure~\ref{figcompl}, the
   R magnitude primarily determines the efficiency in the redshift
   measurements, since most of the unidentified objects are those with
   the reddest colors (i.e., faintest optical magnitudes), representing
   the most challenging targets to be spectroscopically identified. 

   The measurement of the spectral quantities was performed using
   semi-automatic procedures which exploited the IRAF task {\tt splot}:
   first, the continuum was automatically fitted in fixed spectral
   ranges, but the procedures also allow interactive adjustments
   of the continuum level in order to improve the line measurement
   in noisy spectra. Then, the equivalent widths (EWs) and fluxes
   of the most prominent emission lines were measured using both
   a Gaussian fitting algorithm and direct integration of the
   continuum-subtracted line profiles. The fluxes measured with the
   two techniques were in excellent agreement (in more than 90\%
   of the cases the two values agree to within 15\%).
   The equivalent widths are given in the rest-frame, with positive
   or negative values for emission or absorption lines respectively.
   The errors were estimated taking into account
   both the measured rms of the continuum close to the lines and the
   difference between the two measurement techniques. 
   We focused our analyses on equivalent width measurements and continuum
   flux ratios, since the absolute spectrophotometry of our spectra is not 
   fully reliable and not corrected for slit losses.

\begin{figure*}[t]
 \centering
  \resizebox{\hsize}{!}{\includegraphics{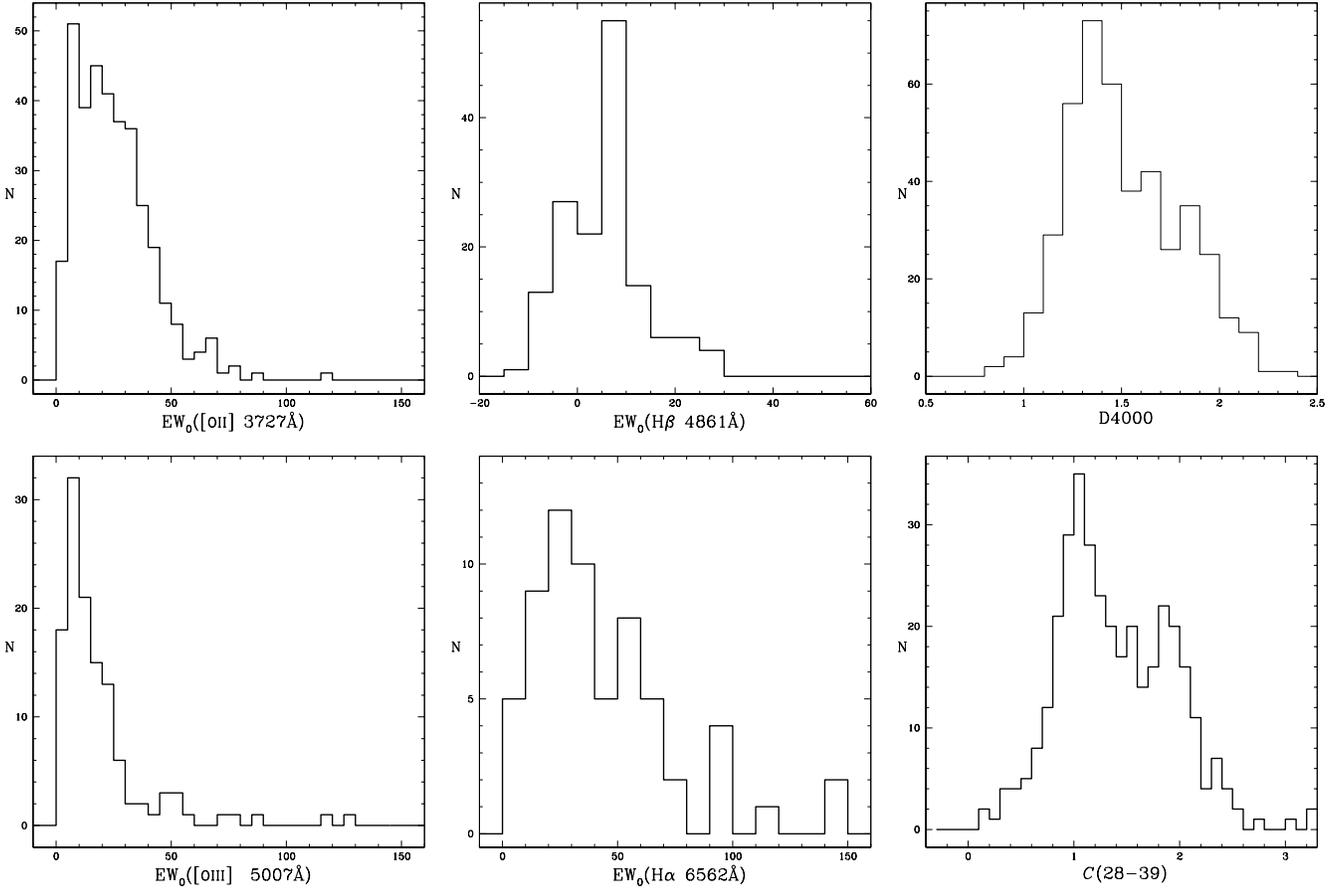}}
    \caption{\footnotesize
     Rest-frame measurement distributions of the main spectral
     features: \OII \ EW (top left panel; 347 measurements),
     \OIII \ EW (bottom left; 122), \Hbeta \ EW (top middle; 148),
     \Halp \ EW (bottom middle; 63), \D4 (top right; 426)
     and our UV color index (bottom right; 349).
     }
   \label{figew}
\end{figure*}
   
   We measured equivalent widths and fluxes for \OII, \Hbeta, \OIII, 
   and \Halp\footnote{our spectral resolution did not allow to separate
   the doublet [N{\tt II}]$\lambda\lambda$6548,6583 from H$\alpha$.},
   as well as the strength of the 4000\AA \ break (\D4;
   Bruzual \cite{Bruzual83}). In addition, an ultra-violet color index
   was calculated by measuring the mean continuum intensity in the
   rest-frame wavelength ranges $2550-3050$~\AA \ 
   (a range typically used to define a star-formation UV indicator;
   see Madau, Pozzetti \& Dickinson \cite{madau}) and $3750-3950$~\AA \
   (the shortward band of the \D4 definition):
 
   \begin{equation}
    \uvcol = -2.5 \log{ \frac {f_\nu(2800)}{f_\nu(3850)}}
    \end{equation}
 
    The continuum flux was measured in fixed spectral ranges,
    and our procedure computes the average with a sigma-clipping
    technique, ensuring that spikes due to bad sky subtraction and/or to 
    cosmic rays residuals did not affect the measured values. 
    The errors were then computed from the standard deviation.
    Figure~\ref{figew} shows the distribution of measurements of the
    main spectral features for the entire sample. The median uncertainty
    for an individual equivalent width is $\sim0.9$ \AA. 
    It is worth noting that we have not accounted for the stellar
    absorption in the measurements of the Balmer emission lines, since
    the spectral resolution and the low S/N in our spectra do not allow
    a simultaneous fit of both the absorption and emission components.
    Usually the effect of the underlying stellar component in absorption
    is accounted for by applying an `ad hoc' correction to the rest-frame EWs
    (see Kennicutt \cite{K92}, Lamareille et~al. \cite{2dFLZ}), but since
    these correction values are determined from local galaxy samples we
    preferred to leave the equivalent widths uncorrected. 
    As seen in the top middle panel of Figure~\ref{figew}, 
    only the \Hbeta \ line is observed both in emission and absorption.
    The absence of H$\alpha$ in absorption is probably due to the
    following causes: first, of the 41 galaxies with \Hbeta \ measured
    in absorption, only 8 include H$\alpha$ in the spectral range, four of
    them with a faint (EW$\sim$4) emission line and four without any
    measurable feature. Moreover, since on average the EW of the
    H$\alpha$~line in absorption is expected to be similar to that of H$\beta$
    (see Charlot \& Longhetti \cite{CeL} and references therein), it is not
    surprising that we did not measure H$\alpha$ in absorption in the
    reddest and usually noisy spectral region.

    The two color index distributions show a bimodal structure,
    more evident in the \uvcol. This property, already seen both
    in the observed color distribution (Balogh et~al. \cite{balogh},
    Bell et~al. \cite{bell04}) and in the intrinsic color indices
    (Kauffmann et~al. \cite{kauff}),
    is a manifestation of underlying blue and red galaxy populations.
    It is worth to note that the \D4 distribution in the K20 survey
    is remarkably similar to those of optical surveys such as
    the Stromlo-APM Survey ($b_j$-selected, Tresse et~al. \cite{tresse})
    and the Sloan Digital Sky Survey ($r$-selected, Kauffmann et~al.
    \cite{kauff}).
    The bimodality in the color indices has been exploited in our endeavor
    to classify the K20 galaxies in different, and coherent, spectral classes
    (see below). 

\begin{figure}[t]
 \centering
  \resizebox{\hsize}{!}{\includegraphics{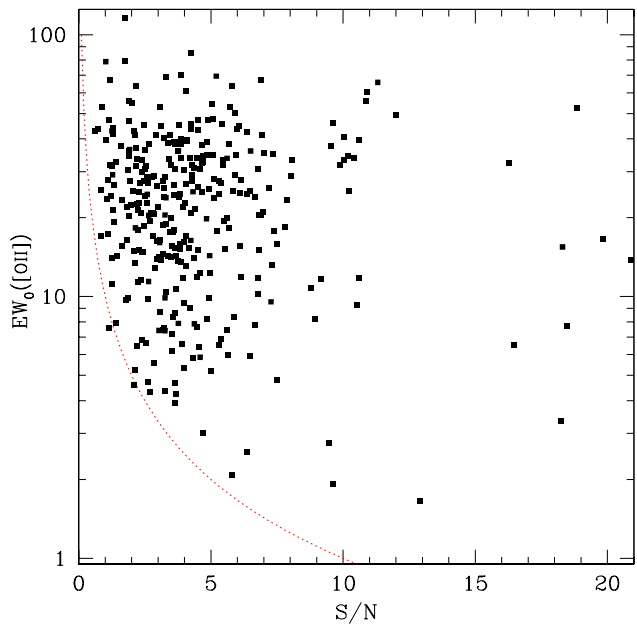}}
    \caption{\footnotesize
     The rest-frame equivalent width of the \OII \ line plotted
     against the signal-to-noise ratio (S/N) of the continuum
     close to the line. The dotted line is the EW detection
     limit curve as estimated by Equation (2), representing a
     good lower envelope of the data.}
   \label{figew_upper}
\end{figure}

    In order to estimate our emission-line detection limits,
    the signal-to-noise ratio in the continuum adjacent to the lines
    was measured.
    Figure~\ref{figew_upper} shows the relation between the S/N in
    the continuum and the rest-frame equivalent widths for the \OII,
    the most common line in our sample: from this figure it is clear that
    the emission line has been detected down to a rest-frame equivalent
    width of $\approx 4$\AA \ for the bulk of the K20 galaxies. The lower
    envelope of the measured EWs (dotted line in Figure~\ref{figew_upper})
    is well represented by the curve

   \begin{equation}
    EW(detection\;limit) = {SL*\Delta\over{(S/N)_{cont}}}
    \end{equation}    
    
    where $\Delta$ is the resolution element (in \AA) of our spectra
    and $SL$ is the significance level 
    of the detectable line, expressed in terms of sigma of the continuum
    noise (Colless et~al. \cite{LDSS}; see also Manning \cite{manning}).
    We adopt a $SL(\sigma)=3$ and, measuring the
    S/N in the continuum of the spectra without a detectable emission line,
    we can estimate the EW upper limits also for these galaxies.

\begin{figure}[t]
 \centering
  \resizebox{\hsize}{!}{\includegraphics{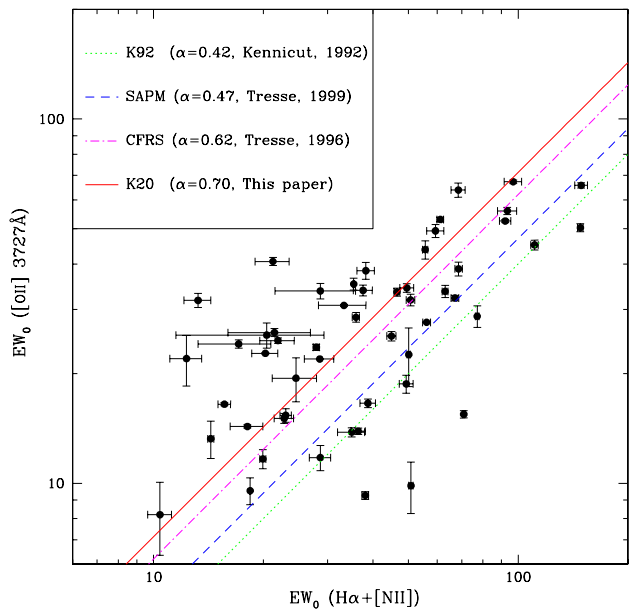}}
    \caption{\footnotesize
     Relationship between the equivalent widths of the \OII \ line and
     \Halp \ complex for the K20 survey. The solid line shows the
     correlation found using our data, while the other lines represent
     the relationships found in the literature [{\it dotted line}, local
     data (Kennicutt \cite{K92}, K92); {\it dashed line}, Stromlo-APM Survey
     (Tresse et~al. \cite{tresse}); {\it dash-dotted line}, Canada-France
     Redshift Survey (Tresse et~al. \cite{CFRSXII})].}
   \label{figOII_Ha}
\end{figure}

    Figure~\ref{figOII_Ha} displays the rest-frame equivalent widths of the
    \OII \ line and \Halp \ complex for the star-forming galaxies
    belonging to the K20 sample:
    the solid line was obtained by fitting the data 
    with an error-weighted linear regression which yields,
    although with a large dispersion (r.m.s. = 21\AA),
    the best-fit relationship
    
   \begin{equation}
    \rm{EW([O{\tt II}]\lambda3727)} = 0.7 \times \rm{EW(H\alpha+[N{\tt II}]\lambda\lambda6548,6583)}.
    \end{equation}
    
    The \OII \ line is often used as an indicator of
    star formation activity since it is the most prominent near-UV line,
    being visible in optical spectra of high-redshift galaxies, where   
    H$\alpha$ becomes inaccessible. A relationship between the equivalent
    widths of the two emission lines was calibrated
    in local spectroscopic surveys (i.e. Kennicutt \cite{K92}, K92).
    In Figure~\ref{figOII_Ha} we show, along with the best-fit to our data
    (solid line), the relationships found in literature,
    such as those obtained by K92 for a local sample of
    galaxies ($z\sim0$, dotted line), by Tresse et~al. (\cite{tresse})
    for the Stromlo-APM survey (SAPM, $<z>\;=0.05$, dashed line) and 
    Tresse et~al. \cite{CFRSXII} for the Canada-France Redshift Survey
    (the CFRS sub-sample with $z<0.3$ and $<z>\,\sim0.2$,
    dash-dotted line). The correlation between $\rm{EW([O{\tt II}]}$)
    and EW(H$\alpha$) in the K20 survey is significantly different from
    that measured locally. Moreover, the various samples suggest
    a trend with redshift, since in our sample the median redshift of
    the galaxies with H$\alpha$ in the observed spectral range is
    $z_{med}=0.29$. In view of these results, the relationship
    for local galaxies calibrated by Kennicutt (\cite{K92}) cannot be
    simply extrapolated to higher redshifts 
    to estimate the star formation rate.
    This is not a new result: it confirms the conclusions drawn by Tresse
    et~al. (\cite{tresse}), but it is important to point out
    that the K20 data confirms the trend to higher redshifts.
    This tendency is also confirmed by near-IR spectroscopy
    of a sample of high-redshift ($z=0.5-1.1$) galaxies from the CFRS
    (Tresse et~al. (\cite{tresse02}).

    These differences in the average ratio between EW(\oii) and
    EW(H$\alpha$) can be ascribed to variations in metallicity or
    ionization state (see Kewley et~al. \cite{kewley}).
    We can confidently exclude that these differences are
    caused by extinction, because we did not find any redshift
    dependence in the extinction indicators we measured
    (i.e., the mean H$\alpha$/H$\beta$ line ratio is similar to the
    value measured in the K92 sample; see Caira et~al., in preparation).
    It is nowadays commonly believed that this effect mainly reflects
    an observational bias introduced by the different luminosity ranges
    bracketed in the various samples rather than a true evolutionary
    redshift trend, since various authors have clearly shown that the
    \oii/H$\alpha$ flux ratio correlates with luminosity both
    in nearby surveys (SAPM survey, Charlot et~al. \cite{Charlot02};
    Nearby Field Galaxy Survey, Jansen, Franx, \& Fabricant \cite{jansen})
    and in high-redshift galaxy samples (Tresse et~al. \cite{tresse02}).

\section{Galaxy spectroscopic classifications}

\begin{figure*}[p]
 \centering
  \resizebox{0.61\hsize}{!}{\includegraphics{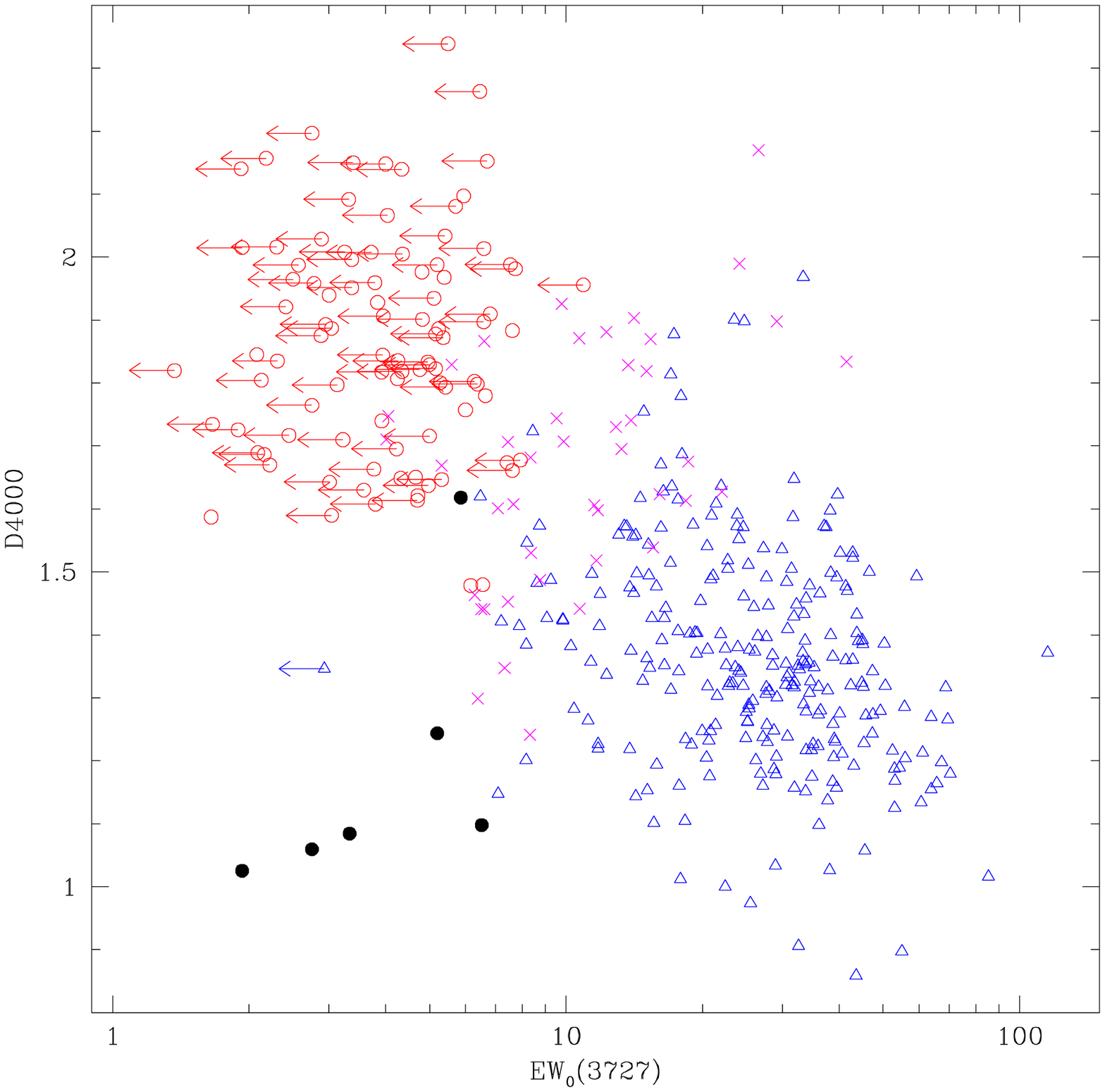}}
    \caption{\footnotesize
    \D4 vs. equivalent width of the \OII \ for the K20 galaxy sample.
    The different symbols refer to the preliminary, visual
    classification: filled circles are type-1 AGN,
    empty circles are early-type passive galaxies (class 1), crosses 
    are red continuum emission-line galaxies (intermediate) and
    empty triangles are blue emission-line galaxies (class 2).
    }
   \label{D4000_ew}

 \centering
  \resizebox{0.61\hsize}{!}{\includegraphics{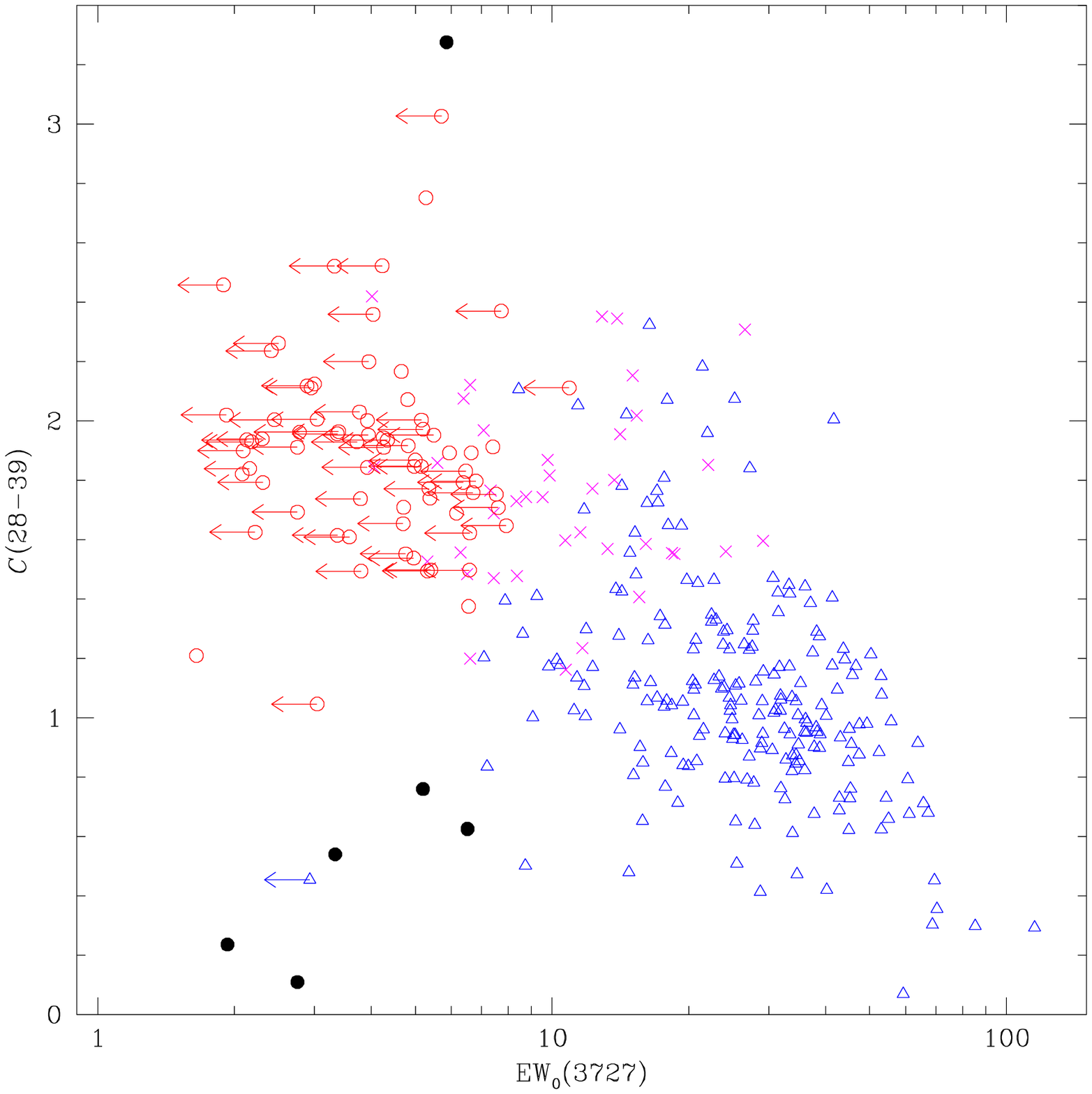}}
    \caption{\footnotesize
    Our UV color index vs equivalent width of the \OII \ for the K20 galaxy
    sample. The different symbols refer to the preliminary, visual
    classification and are the same as in Figure~\ref{D4000_ew}.
    }
   \label{C2839_ew}
\end{figure*}

   A simple \emph{parametric} classification scheme was adopted,
   motivated by the graphs of Figure~\ref{D4000_ew} and
   Figure~\ref{C2839_ew}, where the $\rm{EW_0([O{\tt II}])}$-\D4 and
   $\rm{EW_0([O{\tt II}])}$-\uvcol \ planes are shown, 
   with the different symbols referring
   to the preliminary visual classification: the filled circles are
   type-1 AGN, empty circles represent
   the galaxies with early-type spectra showing no emission lines,
   empty triangles the galaxies with blue emission-line spectra,
   and crosses the intermediate objects, with both a red
   continuum and emission lines. As expected, the continuum indices correlate 
   with the \OII \ equivalent width (and also with the \Halp \ EW, due to
   the relationship between the two lines shown in Figure~\ref{figOII_Ha}).
   Moreover, although the preliminary classification was obviously based on
   visual perception of the same spectral features as those plotted in
   Figures~\ref{D4000_ew} and \ref{C2839_ew}, the segregation of the different
   spectroscopic classes in the EW-continuum index planes is really 
   satisfactory. 
\begin{figure*}[t]
 \centering
  \resizebox{0.9\hsize}{!}{\includegraphics[angle=-90]{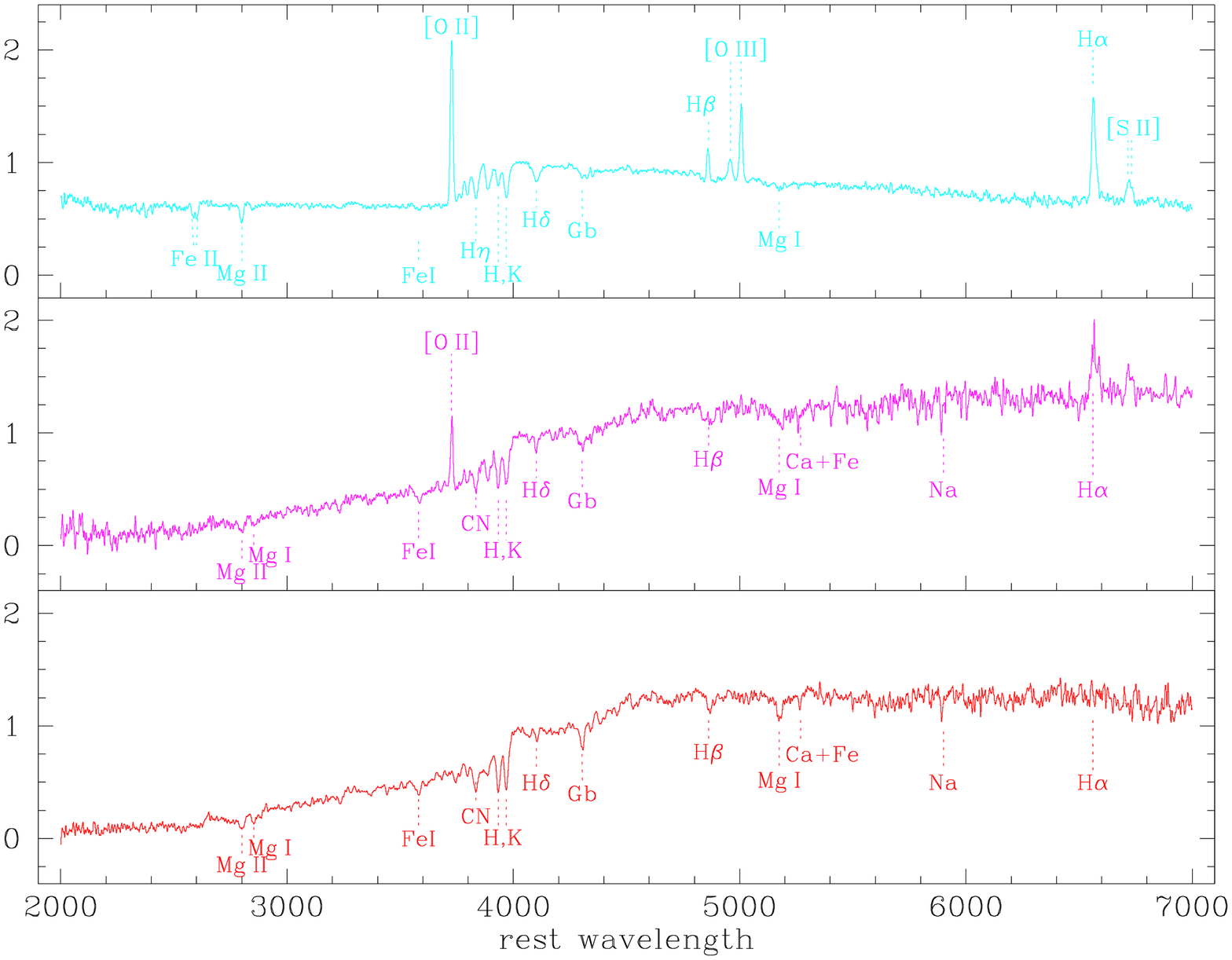}}
    \caption{\footnotesize
    Composite spectra of the three main galaxy classes with the 
    identification of the main absorption and emission lines. From 
    bottom to top: early-type, intermediate and blue
    emission-line galaxy classes. The flux is per unit wavelength
    ($F_\lambda$), and normalization is arbitrary. All the composite
    spectra can be retrieved at the following site: 
    {\tt http://www.arcetri.astro.it/$^\sim$k20/.}}
  \label{plot_ave}
\end{figure*}

   Encouraged by the above findings, we decided to classify the galaxies
   in our sample according to their position in the four-dimensional
   parametric space defined by the EWs of the two main emission lines
   (\OII \ and H$\alpha$+[N{\tt II}]) and by the two continuum indices
   (\D4 and \uvcol).
   We included in our analysis only galaxies with \hbox{$z\le1.6$},
   as the region with the spectral features most useful for
   classification (\D4 \ and EW(\oii)) can be seen only up to this redshift
   in the optical band. In the subsequent discussion, we will use the sample
   of galaxies with secure redshift in the range $0.0 < z \le 1.6$
   (402 objects).
   A four-digit code has been assigned to each galaxy of the sample.
   Each single digit can take one of three values: 0, 1, and 2, if the
   corresponding spectral feature is outside the spectral range, is typical
   of a red passive early-type galaxy or is indicative of a blue active
   emission-line galaxy, respectively. After a careful analysis of
   Figures~\ref{D4000_ew} and \ref{C2839_ew} and after few trials
   in order to minimize the differences between the old and new
   classification systems,
   we fixed the following limits: the leftmost class digit is equal to 1
   if $\rm EW_0($\OII$) < 6 $\AA, the second digit is equal to 1 if
   $\rm EW_0(H\alpha+[N{\tt II}]) < 8 $\AA, the third digit is equal
   to 1 if \D4 $ \ge 1.6 $ and the rightmost digit is equal to 1 if
   \uvcol  $ \ge 1.5 $; each digit value is set to 2 otherwise.
   We consider a galaxy to be classifiable according to this scheme only
   if its code includes at least two non-zero digits. 
   Three main spectroscopic classes have thus been identified:

  \begin {itemize}
   \item \emph{Red, passive early-type galaxies:} all non-zero digits
   are equal to 1. The galaxies belonging to this group are not
   currently undergoing significant star formation detectable with the
   typical sensitivity of our spectra and thus we qualify them as 
   passively evolving.
   \item \emph{Blue, emission-line galaxies:} all non-zero digits
   are equal to 2. These galaxies are undergoing some level  
   of star formation, as witnessed by the relatively intense emission
   lines and blue continuum indices.
   \item \emph{Emission-line galaxies with a red continuum:} the two leftmost
   digits (related to the emission lines) are set to 2, whilst the last
   two digits (continuum indices) are equal to 1. These galaxies show
   an antithetical behavior in the stellar/gaseous components and
   usually present intermediate spectral properties.
  \end {itemize}

    Table~\ref{tabcls} lists the number of K20 galaxies belonging to each
   spectroscopic class, and the composite spectra (see following paragraph)
   of the three main galaxy types are plotted in Figure~\ref{plot_ave}.
   The continuum in the composite of the emission-line galaxies with red
   continuum indices is almost indistinguishable from that of the purely
   passive galaxies composite. Therefore, this class of intermediate objects
   should be mainly composed of elliptical galaxies undergoing a modest
   star-formation episode rather than of heavily reddened star-forming
   galaxies. There are 45 objects with emission lines and red \D4,
   and they represent $\sim$15\% of the whole emission-line galaxy
   population (45/(45+246), see Table~\ref{tabcls}); this percentage is
   in good agreement with the result found in the analysis of the local
   blue-selected SAPM galaxies (see Figure 11 of Tresse et~al. \cite{tresse}).

   The strength of this classification system is highlighted by
   the fact that only 18 out of 402 examined galaxies
   are not included in one of the three main parametric classes.
   These ``outliers'' are composed by a heterogeneous mix of objects,
   such as galaxies with very blue continuum and no emission lines
   (possibly Post-Starburst galaxies), or emission-line galaxies
   with a red \uvcol \ but without a strong \D4, a spectral shape
   that is suggestive of star formation extincted by dust. Indeed,
   to the latter group of galaxies belong most of the dusty starforming 
   EROs identified in the K20 survey (Cimatti et~al. \cite{K20_1}).
   Possibly, some of the unclassified spectra are affected
   by noise or flux calibration problems.
   
   A detailed morphological analysis of the K20 galaxies located
   within the CDFS sky region was performed, exploiting the
   public GOODS HST/ACS imaging available in that field (Cassata
   et~al. \cite{cassata}). The comparison between morphological
   and spectroscopic classification shows an excellent agreement:
   in particular, the objects belonging to the spectral class of
   early-type galaxies are almost all identified with 
   spheroids (elliptical, S0 and bulge-dominated spirals).  
   Galaxies with emission-line spectra appear to correspond
   to a more heterogeneous population of visual morphologies, 
   but the large majority of them ($>$92\%) are associated with
   late-type and irregular ones.
   
   Finally, we did not attempt to sub-divide the galaxies with
   emission lines into Starburst, type-2 Seyfert and LINER classes;
   given the large redshift range observed and the spectral window
   limited to the optical, in the majority of the objects not all
   the diagnostic lines required for an accurate spectral
   classification were present in the spectra. 
%
   \begin{table}[t]
      \caption[]{Number of galaxies belonging to the different spectroscopic
      classes.}
         \label{tabcls}
     $$ 
         \begin{tabular}{lrrr}
            \hline
            \noalign{\smallskip}
            & Total &  CDFS & Q0055 \\
            \noalign{\smallskip}
            \hline
            \noalign{\smallskip}
	    AGN (type~1)                & ~12 & ~~8 & ~~4 \\
	    early-type Galaxies         & ~93 & ~61 & ~32 \\
	    blue Emission-Line Galaxies & 246 & 156 & ~90 \\
	    red Emission-Line Galaxies  & ~45 & ~26 & ~19 \\
	    ``Outliers''                & ~18 & ~13 & ~~5 \\
             \noalign{\smallskip}
            \hline
         \end{tabular}
     $$ 
   \end{table}
%

\section{Composite spectra}

   Two composite spectra were generated for each galaxy class by
   deriving both the mean and median of all of the spectra included 
   in that class. 
   We found that the two resulting composite spectra do not differ
   significantly when at least $\sim$20 spectra were combined.
   Only the brightest emission lines (\oii \ and H$\alpha+[N{\tt II}]$)
   show slightly smaller equivalent widths in the median composite spectra,
   as expected due to the skewed distribution of the EWs
   (see Figure~\ref{figew}).
   In order to build up the composites, each spectrum was 
   shifted to the rest-frame according to its redshift
   (with a 2 \AA \ rest-frame bin\footnote{At the median redshift of the
   analyzed galaxy sample (z=0.73), a rest frame bin of 2 \AA \ is only
   slightly less than the typical pixel size used in the
   spectroscopic observations.}), and normalized in the 4000-4500 \AA \
   wavelength range, which is always present in the observed spectroscopic
   window. To each individual spectrum was assigned the same weight, thus
   avoiding biasing the final composite towards the brightest galaxies.
   Finally, all the spectra belonging to a common spectroscopic class
   were stacked, and along with the average (median) spectrum, also
   a sigma (semi-quartile) spectrum was computed.   

\begin{figure}[t]
 \centering
  \resizebox{\hsize}{!}{\includegraphics{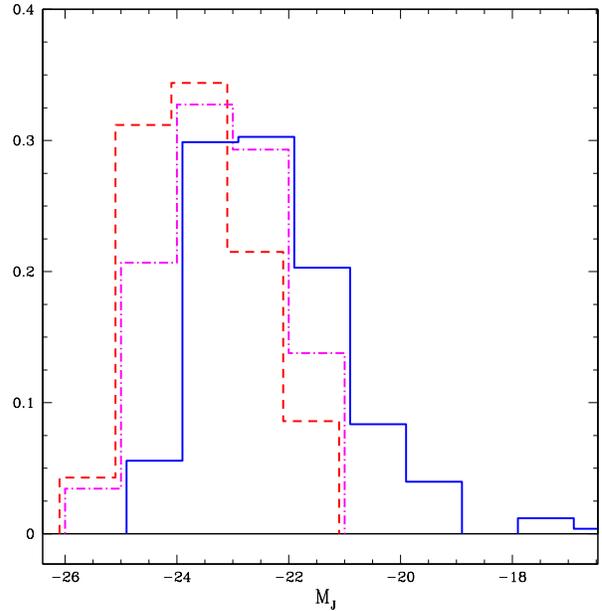}}
    \caption{\footnotesize
     Normalized distributions of the absolute magnitude  computed
     in the rest frame $J$-band ($M_J$) for the different spectroscopic
     classes: the solid histogram is for blue emission-line galaxies,
     the dashed histogram for passive early-type galaxies and the
     dash-dotted histogram for red continuum emission-line galaxies.
    }
  \label{hist_MJ}
\end{figure}

   Hereinafter we adopt the {\it mean\/} composite spectra, but all of the
   following results would apply equally if the median composites were used.
   In Figure~\ref{plot_ave} the average spectra of the three main
   spectroscopic classes are plotted.
   
   Figure~\ref{hist_MJ} shows the frequency distribution, for the different
   spectroscopic classes, of the absolute magnitude computed in the rest
   frame $J$-band following Pozzetti et~al. (\cite{K20_5}, Paper V).
   The distribution of the blue emission-line
   galaxies and that of the early-type galaxies (both passive and with 
   emission lines) are significantly different (the Kolmogorov-Smirnov
   test indicates that the two distributions differ at much more than
   99.9 per cent confidence), red galaxies being, as expected,
   more luminous on average than star-forming ones.
%
%
   \begin{table*}[t]
      \caption[]{Extract of the spectroscopic web catalogue of the Q0055 field.}
         \label{tabcat}
     $$ 
         \begin{tabular}{lllllcccl}
            \hline
            \noalign{\smallskip}
             K20 Id.
            & RA (J2000)
            & DE (J2000)
            & $K$mag
            & $R$-$K^{\mathrm{a}}$
            & redshift
            & $zq^{\mathrm{b}}$
            & cls$^{\mathrm{c}}$
            & spectrum$^{\mathrm{d}}$
	    \\
            \noalign{\smallskip}
            \hline
            \noalign{\smallskip}
q0055\_00013 & 00:57:55.246 & -26:40:57.92 & 18.83 & 3.69 & 1.982 & 1 &  4  & \underline{FITS file}
\\
q0055\_00017 & 00:57:58.218 & -26:40:59.18 & 19.68 & 4.16 & 1.128 & 1 &  2  & \underline{FITS file}
\\
q0055\_00018 & 00:58:02.880 & -26:41:01.06 & 16.16 & 3.00 & 0.200 & 1 &  2  & \underline{FITS file}
\\
q0055\_00019 & 00:58:06.657 & -26:40:59.32 & 18.98 & 3.99 & 0.932 & 0 &  -  & \underline{FITS file}
\\
q0055\_00021 & 00:57:51.516 & -26:41:03.22 & 18.54 & 4.36 & 0.753 & 1 &  1  & \underline{FITS file}
\\
q0055\_00022 & 00:57:59.509 & -26:41:04.21 & 19.19 & 3.01 & 0.679 & 1 &  2  & \underline{FITS file}
\\
q0055\_00023 & 00:57:56.800 & -26:41:04.73 & 19.87 & 3.19 & 1.436 & 1 &  2  & \underline{FITS file}
\\
q0055\_00024 & 00:57:54.719 & -26:41:07.51 & 19.94 & 2.39 & 1.374 & 1 &  2  & \underline{FITS file}
\\
q0055\_00025 & 00:57:54.499 & -26:41:09.32 & 18.46 & 3.64 & 0.823 & 1 &  2  & \underline{FITS file}
\\
q0055\_00026 & 00:57:52.065 & -26:41:13.80 & 17.25 & 4.34 & 0.794 & 1 &  3  & \underline{FITS file}
\\
q0055\_00028 & 00:57:50.333 & -26:41:09.29 & 19.10 & 5.09 & 1.050 & 1 &  1  & \underline{FITS file}
\\
q0055\_00029 & 00:57:53.405 & -26:41:14.90 & 19.86 & 2.64 & 0.416 & 1 &  2  & \underline{FITS file}
\\
q0055\_00030 & 00:57:56.601 & -26:41:14.66 & 19.50 & 3.62 & 0.565 & 1 &  2  & \underline{FITS file}
\\
q0055\_00031 & 00:58:06.350 & -26:41:15.63 & 17.98 & 4.80 & 0.679 & 1 & 1.5 & \underline{FITS file}
\\
q0055\_00033 & 00:57:56.171 & -26:41:15.73 & 18.98 & 5.47 & 0.820 & 1 &  2  & \underline{FITS file}
\\
q0055\_00034 & 00:58:02.447 & -26:41:16.12 & 18.59 & 4.34 & 1.129 & 1 &  2  & \underline{FITS file}
\\
q0055\_00035 & 00:58:03.507 & -26:41:15.89 & 18.84 & 3.61 & 0.668 & 1 &  2  & \underline{FITS file}
\\
q0055\_00036 & 00:57:51.544 & -26:41:16.75 & 19.53 & 3.52 & 0.925 & 1 &  2  & \underline{FITS file}
\\
q0055\_00038 & 00:58:04.790 & -26:41:18.90 & 18.57 & 4.32 & 0.805 & 1 &  3  & \underline{FITS file}
\\
q0055\_00039 & 00:57:56.361 & -26:41:18.92 & 18.67 & 3.78 & 0.561 & 1 & 1.5 & \underline{FITS file}
\\
q0055\_00040 & 00:57:57.748 & -26:41:19.07 & 18.55 & 5.47 & 1.210 & 1 & 1.5 & \underline{FITS file}
\\
q0055\_00041 & 00:58:01.231 & -26:41:19.98 & 18.70 & 4.66 & 0.921 & 1 & 1.5 & \underline{FITS file}
\\
q0055\_00043 & 00:57:51.685 & -26:41:22.30 & 19.20 & 2.68 & 0.580 & 1 &  2  & \underline{FITS file}
\\
q0055\_00045 & 00:58:00.764 & -26:41:24.18 & 18.99 & 6.87 &   -   & - &  -  & NOT OBSERVED
\\
q0055\_00046 & 00:57:49.190 & -26:41:27.01 & 16.61 & 4.35 & 0.565 & 1 &  1  & \underline{FITS file}
\\
q0055\_00047 & 00:57:57.074 & -26:41:30.03 & 14.64 & 3.55 & 0.266 & 1 &  1  & \underline{FITS file}
\\
q0055\_00049 & 00:57:54.200 & -26:41:27.79 & 16.63 & 3.81 & 0.265 & 1 & 1.5 & \underline{FITS file}
\\
q0055\_00050 & 00:58:05.671 & -26:41:26.96 & 19.25 & 4.41 & 1.002 & 1 &  2  & \underline{FITS file}
\\
q0055\_00052 & 00:58:07.025 & -26:41:27.04 & 18.99 & 2.89 & 2.175 & 1 &  -  & \underline{FITS file}
\\
q0055\_00053 & 00:58:07.960 & -26:41:27.98 & 19.53 & 4.95 & INDEF & -1 &  -  & \underline{FITS file}
\\
q0055\_00056 & 00:58:01.435 & -26:41:31.96 & 17.72 & 3.13 & 0.431 & 1 &  2  & \underline{FITS file}
\\
q0055\_00059 & 00:58:04.129 & -26:41:34.42 & 19.25 & 4.72 & INDEF & -1 &  -  & \underline{FITS file}
\\
................... & .................... & .................. & ........ & ....... & ..... & ... & ... & ..............
\\
\noalign{\smallskip}
\hline
\end{tabular}
     $$ 
\begin{list}{}{}
\item[($^{\mathrm{a}}$)] $R$-$K$ color, measured in a 2\arcsec diameter aperture.
\item[($^{\mathrm{b}}$)] $zq$ stands for `z-quality', 
the reliability of the redshift determination.
\item[($^{\mathrm{c}}$)] spectroscopic class (see text).
\item[($^{\mathrm{d}}$)] hyperlink to the spectrum FITS file.
\end{list}
   \end{table*}

\section{Release of the spectroscopic catalogue}

   An excerpt from the spectroscopic catalogue of the Q0055 field is
   presented in Table~\ref{tabcat}; the full catalogues (for both the
   CDFS and Q0055 fields) are available at the K20 web site
   {\tt http://www.arcetri.astro.it/$^\sim$k20/}. In detail the content
   of Table~\ref{tabcat} is as follows:
 \begin {itemize}
   \item \emph{Column 1:} Internal K20 identification number;
   \item \emph{Column 2-3:} Right ascension and declination (equinox 2000.0);
   \item \emph{Column 4:} Total (SExtractor BEST) $K_s$-band magnitude;
   \item \emph{Column 5:} $R$-$K$ color, measured in a 2\arcsec-diameter aperture;
   \item \emph{Column 6:} Redshift;
   \item \emph{Column 7:} Redshift quality flag: ``1'' indicates a solid
   redshift determination, ``0'' a tentative redshift determination and ``-1''
   no redshift determination;
   \item \emph{Column 8:} Spectroscopic classes as assigned in Section 5:
   ``0'' is for objects classified as stars, ``1'' is for red passive
   early-type galaxies (four-digit classification = 1111), ``2'' for blue
   emission-line galaxies (2222), ``1.5'' for intermediate galaxies with
   emission lines but red continuum indices (2211), ''3'' for galaxies which
   are not included in one of the three previous classes and ''4'' is for
   broad-line AGN. We recall that objects with only a tentative redshift
   ($zq=0$) are not classified.
   \item \emph{Column 9:} In the web catalogue the last column contains a
   hyperlink to the one-dimensional object spectrum, which can be downloaded
   as a FITS file.
  \end {itemize}   

\section{Evolution of the spectral properties of the K20 spectroscopic sample}

   The majority of the spectra in the K20 survey was of relatively low $S/N$
   and consequently the measurement of the individual spectral parameters was
   subject to substantial uncertainties. Therefore we decided to explore
   the average spectral evolution of different populations using the
   higher $S/N$ composite spectra of galaxy classes.
   The average spectra of spectroscopic subsamples
   selected by redshift and/or luminosity are known to be well suited
   to study the mean properties of galaxy populations, and the following
   discussion is based mainly upon such composites. An analysis of
   individual object spectra can be found in Caira (\cite{caira})
   and Caira et~al., in preparation.

\begin{figure*}[t]
 \centering
  \resizebox{0.9\hsize}{!}{\includegraphics[angle=-90]{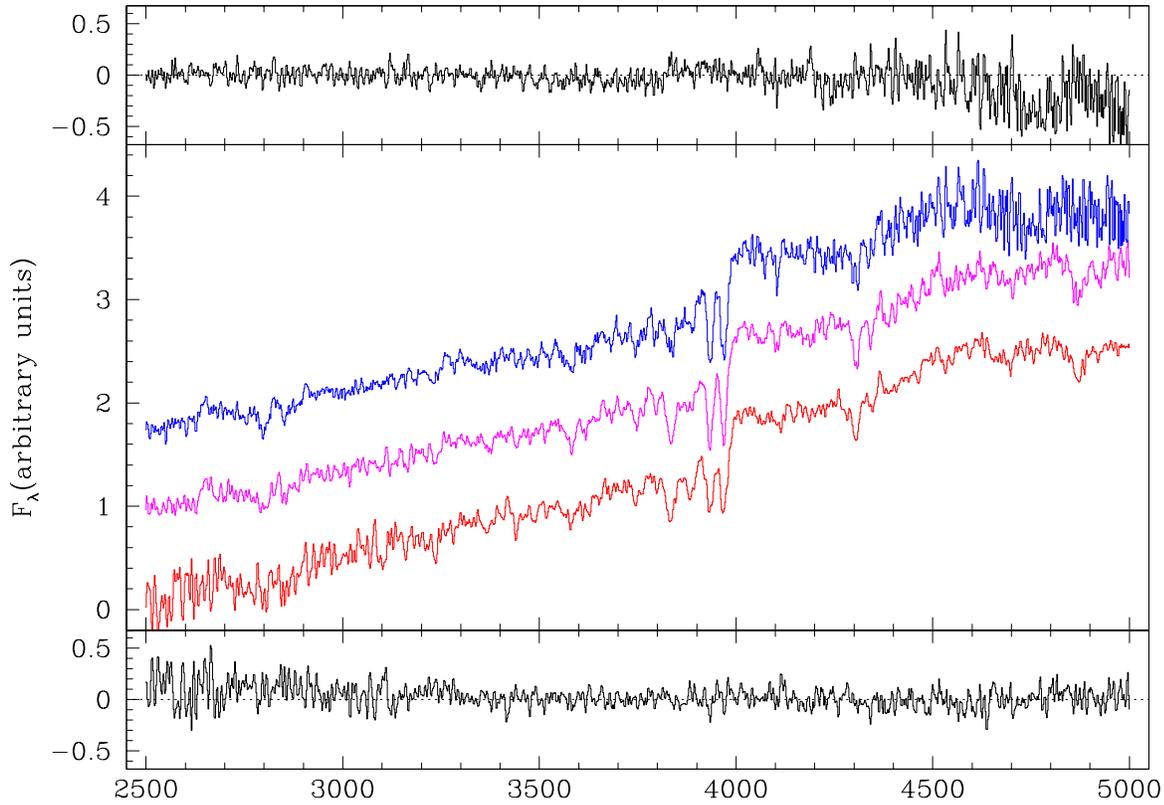}}
    \caption{\footnotesize
    Composite spectra of the early-type galaxies divided in
    three redshift bins. From bottom to top: the composite of 18 low-z
    ($0 < z < 0.6$) galaxies, the composite of 49 intermediate-z
    ($0.6 < z < 0.75$) galaxies and the composite of 26 high-z
    ($0.75 < z < 1.25$) galaxies. Almost all the galaxies included
    in the intermediate-z sample belong to the two overdensities present
    in the K20 survey area. The flux normalization is arbitrary
    and the spectra are offset for clarity. The upper and lower panels
    also show, with the same scale, the difference spectra between the
    intermediate-redshift range composite and, respectively,  
    the high-z composite and the low-z composite spectra.}
  \label{plot_ave_Early}
\end{figure*}

\subsection{Early-type galaxies}
   Figure~\ref{plot_ave_Early} shows the composite spectra of the K20
   objects belonging to the class of the early-type galaxies, divided in
   three redshift bins: in the low-z interval \hbox{($0<z<0.6$)} 18~objects
   are included, and the median redshift is $z_{med}=0.52$; 
   in the high-z interval \hbox{($0.75<z<1.25$)} 26~objects are included,
   with $z_{med}=1.02$. 
   The highest number of galaxies (49) is included in the
   intermediate-redshift range ($0.6<z<0.75$), with $z_{med}=0.68$ 
   and with the majority of the objects belonging to the two overdensities
   which are present in the K20 survey area (see Paper III).
   The three composite spectra are remarkably similar, as clearly shown by
   the upper/lower panels of Fig.~\ref{plot_ave_Early}, where
   the difference spectra between high-z and intermediate-z composites (top),
   and between low-z and intermediate-z composites (bottom) are displayed.
   The difference spectra are smooth, with zero mean and
   the relative variation\footnote{$(spe1-spe2)/<spe1,spe2>$} between two
   composite spectra in contiguous redshift bins is typically smaller than
   6\% over the full wavelength range.
   The deviations found at the edges of the difference spectra are probably
   due to the noise increasing in the blue part of the low-z composite
   and in the red part of the high-z composite, since the number
   of spectra contributing to the mean is lower in these wavelength ranges.

\begin{figure}[t]
 \centering
  \resizebox{\hsize}{!}{\includegraphics{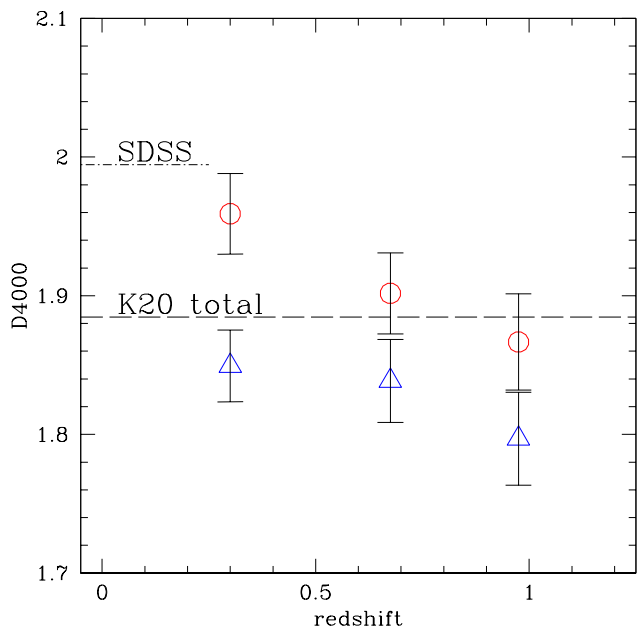}}
    \caption{\footnotesize \D4 strengths as a function of redshift in K20
    early-type composites. The galaxies in each redshift bin have been
    divided in two equally populated groups, according to
    their luminosity, and circles and triangles indicate the brighter
    and fainter subsamples, respectively. The dashed line indicates
    the \D4 value measured in the average spectrum of the whole early-type
    class of Fig.~\ref{plot_ave}, whereas the dash-dotted line is
    the \D4 measured in the SDSS composite of early-type galaxies
    (MAIN sample, Eisenstein et~al. \cite{SDSS_E}). 
  }
  \label{D4_z}
\end{figure}

   The average composites of the early-type class are also
   strikingly similar to the Sloan Digital Sky Survey (SDSS) mean
   spectrum for the same galaxy population (Eisenstein et al. \cite{SDSS_E}),
   although small  differences are noted in the 4000\AA \ break and
   continuum colors, which are systematically bluer in our templates.
   We decided to further explore the evolution of the K20 early galaxy 
   population by checking the luminosity dependence of the
   composites: in each redshift bin we divided the galaxies in two equally
   populated groups, according to their absolute magnitude M$_J$, and
   we created the corresponding composites. Figure~\ref{D4_z} shows the
   strengths of the \D4 index measured in the six average spectra,
   where the circles and the triangles represent the brighter and fainter
   galaxy populations, respectively. It is evident that small
   (note the different scale of the y-axes in Figure~\ref{D4_z}
   and~\ref{D4000_ew}) but systematic differences are present in the
   \D4 index. This plot shows a correlation of the \D4 index with both
   redshift and luminosity; qualitatively, the trend with redshift is 
   consistent with an older stellar population at lower redshift
   (at fixed metallicity), while the trend with luminosity is
   consistent with higher formation redshift for the more
   luminous early-type galaxies.
   
   Such mild evolution up to $z \sim 1$ in the early-type galaxy
   population, at least in the mean spectral characteristics, is in
   agreement with the previous results of the K20 survey, namely:
   1) the high-redshift formation of the ``old'' passively evolving EROs
   (Cimatti et~al. \cite{K20_1}, Paper I); 
   2) the small, if any, decrease of the number density of the early-type
   galaxies up to $z\simeq 1$ (Pozzetti et~al. \cite{K20_5}, Paper V);
   3) the existence of old, massive spheroidal galaxies at $z > 1.5$
   (Cimatti et~al. \cite{oldyoung}) and 
   4) the change, with redshift and stellar mass content,
   of the M$_{*}$/L ratio in early-type galaxies, suggesting
   that the stellar population formed at higher redshift in the most
   massive early-type galaxies (Fontana et~al. \cite{K20_6}, Paper VI). 
   All the above statements fit into a coherent picture where a substantial
   number of massive spheroids were fully assembled at redshifts much
   greater than one, an essential piece of information for the understanding
   of the formation and evolution of elliptical galaxies (see also
   Glazebrook et~al. \cite{GDDS}).

\subsection{Blue emission-line galaxies}
\begin{figure*}[t]
 \centering
  \resizebox{0.9\hsize}{!}{\includegraphics[angle=-90]{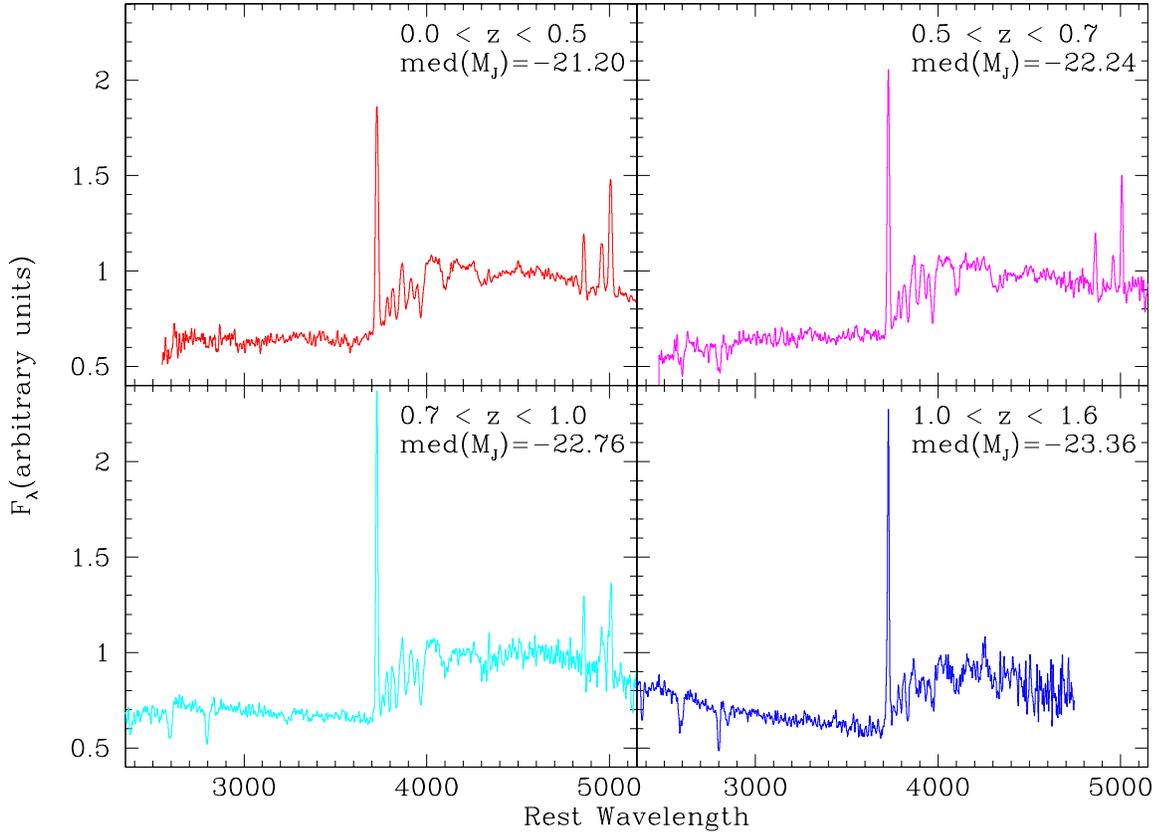}}
    \caption{\footnotesize
    Composite spectra of the $M_J<-20$ blue emission-line galaxies in
    four redshift bins. The flux normalization is arbitrary and the
    spectra are shown only in the wavelength range in which more than
    20 spectra contribute to the template. The apparent intensity 
    increase of the \emph{unresolved} \oii \ emission line with
    redshift is a visual effect due to the enhancement of rest-frame
    spectral resolution in the high-z composites.
    }
  \label{4spe}
\end{figure*}

    A different behavior is shown by the average spectra of the
   class of emission-line blue galaxies. The composite spectra
   in four redshift intervals are presented in Figure~\ref{4spe}.
   The redshift intervals have been chosen in such a way to have
   a similar number of objects in each bin, and the galaxies with
   absolute magnitude $M_j>-20$, which are observed only locally
   ($z<0.25$), have been excluded; the composite spectra
   are shown only in the wavelength range in which more than
   20 spectra contribute to them. Although galaxies with significantly
   different star-formation histories are likely to contribute to each
   composite spectrum, an analysis of these spectra may still give
   some clues on the average properties of our emission-line
   galaxies as a function of redshift. The shape of the continuum
   shows a blueing with redshift, particularly evident in the highest
   redshift bin, where the spectrum increases toward the blue
   at wavelengths shortward than \oii. This is quantitatively
   seen in Table~\ref{tabparz}, where 
   the number of objects contributing to the four composite spectra
   shown in Fig.~\ref{4spe}, their median absolute magnitude $M_j$
   and the measured value of EW(\oii), \D4 and \uvcol \ are given.
   While the EW(\oii) is approximately constant in the four spectra,
   the value of \uvcol \ in the two high-redshift bins is significantly
   smaller than the value measured in the low redshift bins.

   In order to explain the observed redshift evolution, the four
   composite spectra of Fig.~\ref{4spe} have been fitted
   (continuum + absorption lines) using the BC03 models
   (Bruzual \& Charlot \cite{BC03})
   with exponentially decreasing star formation with time-scale $\tau$,
   a Salpeter IMF with M$_{up}$ = 125~M$_{\sun}$, 
   solar metallicity, and with ages $t$ running 
   up to the age of the Universe at the corresponding average redshifts.
   In addition, in absence of reliable indicators of extinction
   (only for a sub-sample of the galaxies in the first redshift
   bin do we have H$\alpha$ in our spectra), we have adopted models with
   no extinction. As a consequence, the results we briefly discuss
   below should be taken only as indicative of possible trends.
   The spectra in the first two redshift bins are well represented
   by models with $t/\tau \sim 2.0-2.5$, 
   which correspond to a ratio between the current SFR
   and the past average SFR (the Scalo $b$ parameter; Scalo \cite{scalo})
   in the interval $0.2-0.3$.
   With the same model, the composite spectrum in the highest
   redshift bin is better represented with $t/\tau \sim 1.2-1.8$
   and the corresponding $b$ values are in the interval $0.4-0.5$.
%
   \begin{table*}[t]
      \caption[]{Spectral Measurements in Emission Line Composite as a function
      of redshift.}
         \label{tabparz}
     $$ 
         \begin{tabular}{lcclccc}
            \hline
            \noalign{\smallskip}
            z bin &  sample & N$_{obj}$ & med(M$_J$) & EW$_0$(\oii) & \D4 & \uvcol \\
            \noalign{\smallskip}
            \hline\hline
            \noalign{\smallskip}
$ 0.0\div0.5 $&$     M_J\leq-20     $&$ 51 $&$ -21.20 $&$ 30.9\pm0.6 $&$ 1.37\pm0.02 $&$ 1.05\pm0.03 $\\
$ 0.5\div0.7 $&$     M_J\leq-20     $&$ 60 $&$ -22.24 $&$ 29.3\pm0.6 $&$ 1.34\pm0.02 $&$ 1.11\pm0.03 $\\
$ 0.7\div1.0 $&$     M_J\leq-20     $&$ 64 $&$ -22.76 $&$ 33.7\pm0.3 $&$ 1.30\pm0.02 $&$ 0.93\pm0.03 $\\
$ 1.0\div1.6 $&$     M_J\leq-20     $&$ 53 $&$ -23.38 $&$ 31.4\pm0.4 $&$ 1.30\pm0.02 $&$ 0.85\pm0.03 $\\
            \noalign{\smallskip}
            \hline\hline
         \end{tabular}
     $$ 
   \end{table*}

   These fits suggest that, as expected, the high-redshift emission-line
   galaxies are ``younger'' and more ``active'' (i.e., with higher $b$)
   than those detected at lower redshift. For a given model, with the
   same mass function, metallicity and ionization parameter ($U$),
   we would therefore expect a higher \oii \ equivalent width for
   the high-redshift galaxies (Magris et~al. \cite{magris}), while the
   measured equivalent widths in the four composites are similar to
   each other (see Table~\ref{tabparz}). The observed values of the
   \oii~EW ($\sim$30\AA) suggest that the average metallicity
   of the galaxies in the four composite spectra are solar or sub-solar
   (see Figure 8 in Magris et~al. \cite{magris}).
   From the dust-free models computed by Magris
   et~al.\footnote{See http://www.cida.ve/$^\sim$magris/ademis/datafile1.txt.},
   we find that the expected \oii \ equivalent width for the best-fit
   models of the high-redshift composite is indeed of the order of $25-30\,$\AA \
   for $log\;U = -3.00 \pm 0.25$, while they are of the order of $20-22\,$\AA \
   for the low-redshift composite. The higher observed equivalent width
   would apparently require a somewhat lower average metallicity for the
   low-redshift objects. Since the low-luminosity galaxies 
   are included only in the lower redshift bins, this effect might be induced
   by the luminosity-metallicity correlation for star forming galaxies
   (see Kobulnicky \cite{kobul} for a recent review of its
   evolution with redshift). To better evaluate the possibility of a
   luminosity effect in our composite spectra, in each redshift bin
   we have further subdivided the galaxies in two equally populated
   sub-samples ($\sim 25-30$ objects in each sub-sample) as a function of
   absolute magnitude M$_J$. Indeed, we find that the average spectra of
   the faint galaxies are bluer in all redshift bins, with higher
   \oii \ equivalent widths, except for the highest redshift
   bin in which the two values for the bright and faint sub-sample
   are consistent with each other. 
   In addition, in the low redshift composites the
   [O{\tt III}]/H$\beta$ ratio is significantly higher in the sub-sample
   of fainter galaxies, consistently with the expectations from a
   metallicity - luminosity correlation. Note, however, that these
   differences in the [O{\tt III}]/H$\beta$ ratio can also be due to
   different average values of the ionization parameter $U$.
   A more detailed analysis of the properties of the star-forming galaxies
   in our sample will be presented elsewhere (Caira et~al., in preparation).

\section{Conclusions}

 In this paper we presented a spectroscopic study of galaxies
 selected in the K20 survey, based on the optical data obtained at the
 ESO VLT with the FORS1 and FORS2 spectrographs. The K20 sample includes
 545 objects down to $K_s = 20.0$, and we observed 525 of them,
 providing 501 redshift identifications. The spectroscopic completeness
 is therefore 92\%, the highest to date for surveys selected in the
 near-infrared. The identified sample includes 45 stars, 12 broad-line
 AGN and 444 ``normal'' galaxies, 412 of which have highly reliable
 redshifts. The spectral catalogue, with derived redshifts and reduced
 spectra, has been released to the community
 ({\tt http://www.arcetri.astro.it/$^\sim$k20/}).
 
 The main results of this work can be summarized as follows:
 
   \begin{enumerate}
      \item The spectroscopic catalogue of the K20 survey, now released,
      includes the following set of measurements: 
      redshifts, equivalent widths of four optical
      lines (from \oii \ to H$\alpha$), two continuum indices
      (\D4 and \uvcol) and a spectral classification. This large
      collection of information, along with the high completeness of
      the spectroscopic sample, provides a powerful tool
      to investigate the nature and evolution of the extragalactic
      population.
      \item We defined a ``fair'' spectroscopic classification, which
      makes use of four spectral parameters to objectively divide the
      galaxies in three main classes; all but 18 of the K20
      galaxies can be placed in one of the following categories:
      1) early-type galaxies, with only absorbing lines and red 
      continuum; 2) star-forming galaxies, with emission lines and
      blue continuum; 3) intermediate galaxies, with emission lines
      but red continuum, similar to that of the objects belonging
      to the first group.
      \item Composite spectra for the three main spectroscopic classes
      have been computed and used to improve the redshift determination.
      These templates are publicly released along with the
      spectroscopic catalogue.  
      \item Using the composite spectra divided into redshift bins, 
      we found that the average spectra of the early-type galaxies
      are remarkable in their similarity, and only small but 
      systematic differences with redshift are detected in the \D4;
      qualitatively such differences are consistent with the ageing
      of the stellar population.
      Conversely, the star-forming galaxies included in the class
      of objects with emission lines and blue continuum indices
      show a clear and monotonic ``blueing'' of the continuum
      with increasing redshift. We compared the emission line
      galaxy composites with simple dust-free models from BC03,
      finding that the galaxies at high redshift are more 
      active (i.e., have a higher ratio between the current and the
      past average SFR) than those included in the lower redshift bins.
      The observed values of the \oii \ equivalent width
      also suggest a solar or sub-solar metallicity of the galaxies, 
      with a somewhat lower average metallicity for the low-$z$
      objects in order to explain the approximate constancy of the
      \oii~EW throughout the redshift intervals. This may be
      at least partly due to the metallicity-luminosity relationship
      for star-forming galaxies. However, we stress that at this stage
      these interpretations should be taken only as indicative of
      possible trends, as they strongly depend on the assumptions
      made about the extinction and ionization.

   \end{enumerate}

  Most the aspects of the spectroscopic analysis of faint 
  galaxy samples presented in this paper will be investigated 
  further by the current and future massive spectroscopic
  survey, (i.e. SDSS; VIMOS VLT Deep Survey, Le~F\'evre et~al.
  \cite{VVDS}), even though none of these giant projects owns,
  to our knowledge, the dual characteristic of selection in the
  near-infrared band and high level of spectroscopic completeness 
  which make the K20 survey such a powerful instrument for
  the study of galaxy evolution.

\begin{acknowledgements}
   We thank the VLT support astronomers for their kind assistance
 and competent support during the observations. We are grateful
 to C.~Vignali and G.M.~Stirpe for his careful reading of the paper
 and P.~Ciliegi for his help with Figure~\ref{figk20_hudf}.
\end{acknowledgements}

\end{document}